\documentclass[prl,preprint,superscriptaddress]{revtex4}
\usepackage{graphicx}
\usepackage{dcolumn}
\usepackage{amssymb}
\usepackage{bm}
\usepackage{pifont}
\usepackage{epsfig}
\usepackage{psfrag}
\usepackage{epstopdf}
\usepackage{wasysym}
\usepackage[usenames]{color}
\begin{document}

\title{Strength of Mechanical Memories is Maximal at the Yield Point of a Soft Glass}
\author{Srimayee Mukherji}
\affiliation{Chemistry and Physics of Materials Unit, Jawaharlal Nehru Centre for Advanced Scientific Research, Jakkur, Bangalore - 560064, INDIA}
\author{Neelima Kandula}
\affiliation{Chemistry and Physics of Materials Unit, Jawaharlal Nehru Centre for Advanced Scientific Research, Jakkur, Bangalore - 560064, INDIA}
\author{A K Sood}
\affiliation{Department of Physics, Indian Institute of Science, Bangalore - 560012, INDIA}
\affiliation{International Centre for Materials Science, Jawaharlal Nehru Centre for Advanced Scientific Research, Jakkur, Bangalore - 560064, INDIA}
\author{Rajesh Ganapathy}
\affiliation{International Centre for Materials Science, Jawaharlal Nehru Centre for Advanced Scientific Research, Jakkur, Bangalore - 560064, INDIA}
\affiliation{School of Advanced Materials (SAMat), Jawaharlal Nehru Centre for Advanced Scientific Research, Jakkur, Bangalore - 560064, INDIA}
\date{\today}

\draft
\begin{abstract}

We show experimentally that both single and multiple mechanical memories can be encoded in an amorphous bubble raft, a prototypical soft glass, subject to an oscillatory strain. In line with recent numerical results, we find that multiple memories can be formed sans external noise. By systematically investigating memory formation for a range of training strain amplitudes spanning yield, we find clear signatures of memory even beyond yielding. Most strikingly, the extent to which the system recollects memory is largest for training amplitudes near the yield strain and is a direct consequence of the spatial extent over which the system reorganizes during the encoding process. Our study further suggests that the evolution of force networks on training plays a decisive role in memory formation in jammed packings.

\end{abstract}
\maketitle

In a seminal experiment, Paulsen and coworkers observed that the addition of noise helped form memories of multiple strain amplitudes in a periodically sheared dilute non-Brownian suspension \cite{paulsen2014multiple}. This phenomenon, first predicted numerically \cite{keim2011generic}, shares striking similarities with findings on charge density wave solids \cite{povinelli1999noise} and helps distinguish this class of memory from other well-known memory effects \cite{jonason1998memory,sethna1993hysteresis}. Subsequently, simulations found encoding of multiple mechanical memories possible even in amorphous solids subject to a cyclic shear albeit here, the source of noise is intrinsic and stems from the complexity of the energy landscape \cite{fiocco2014encoding}. The ability to form such memories highlights the complex interplay between noise and the underlying reversibility-irreversibility transition (RIT) at a threshold strain amplitude $\gamma_c$ in these systems \cite{pine2005chaos,corte2008random,fiocco2013oscillatory}. The yielding transition in amorphous solids, under oscillatory shear, shares qualitative features with RIT and $\gamma_c$ has been identified with the yield strain $\gamma_y$ \cite{fiocco2013oscillatory,regev2013onset,nagamanasa2014experimental,regev2015reversibility}. In the absence of noise, repeated cycles of training at an amplitude $\gamma_t < \gamma_c$ eventually results in the system reaching a reversible steady state. Since reversibility at $\gamma_t$ implies reversibility for all $\gamma_{\circ} < \gamma_t$, even when trained at multiple amplitudes $\gamma_1,\gamma_2,...,\gamma_n$, with $\gamma_{1}<...<\gamma_n<\gamma_t$, memory of only $\gamma_t$ is retained in the steady state. With noise present, the system meanders around a subset of accessible metastable states and this allows encoding of multiple memories \cite{keim2011generic,povinelli1999noise}. Memory of $\gamma_t$(s) can be retrieved by performing a strain sweep and, if present, manifests itself as cusp(s) in irreversibility as $\gamma_{\circ}$ is swept past $\gamma_t$(s). Even while one expects at least a partial retention of the training in the fluctuating steady state \cite{keim2013multiple}, evidence for memory for $\gamma_t>\gamma_y$ is currently lacking even in simulations on amorphous solids. Probing memory effects across the yielding transition in soft glasses becomes particularly relevant given that local plastic rearrangements in these systems are known to be correlated through long-range elastic interactions \cite{goyon2008spatial,katgert2010couette,nagamanasa2014experimental,schall2007structural,chikkadi2011long,ghosh2017direct}. This is quite unlike sheared dilute suspensions, where particle reorganization events are purely local \cite{corte2008random}. Nevertheless, experiments are yet to find signatures of even single memories, let alone multiple ones, and that too below $\gamma_y$ in amorphous solids. 
  
In this Letter, we explore the formation of mechanical memories in athermal amorphous bubble rafts subject to a cyclic shear. Owing to the negligible friction between bubbles and the qualitative similarities in nature of inter-particle interactions with atomic systems \cite{shi1982potential}, bubble rafts are often considered a champion model system for studying deformation mechanisms of crystals and glasses \cite{shi1982potential,bragg1947dynamical,debregeas2001deformation,lauridsen2004velocity,katgert2010couette,schall2010shear}. We provide the first experimental evidence of both single and multiple memories in a soft glass and more importantly, we observe clear signatures of memory even for $\gamma_t>\gamma_y$. We find that the degree to which the system recollects memory of the training is maximal for $\gamma_t\approx\gamma_y$ and is a direct consequence of the large scale spatial reorganization of the system that occurs for training amplitudes in the vicinity of yield.

\begin{figure}[tbp]
\includegraphics[width=0.8\textwidth]{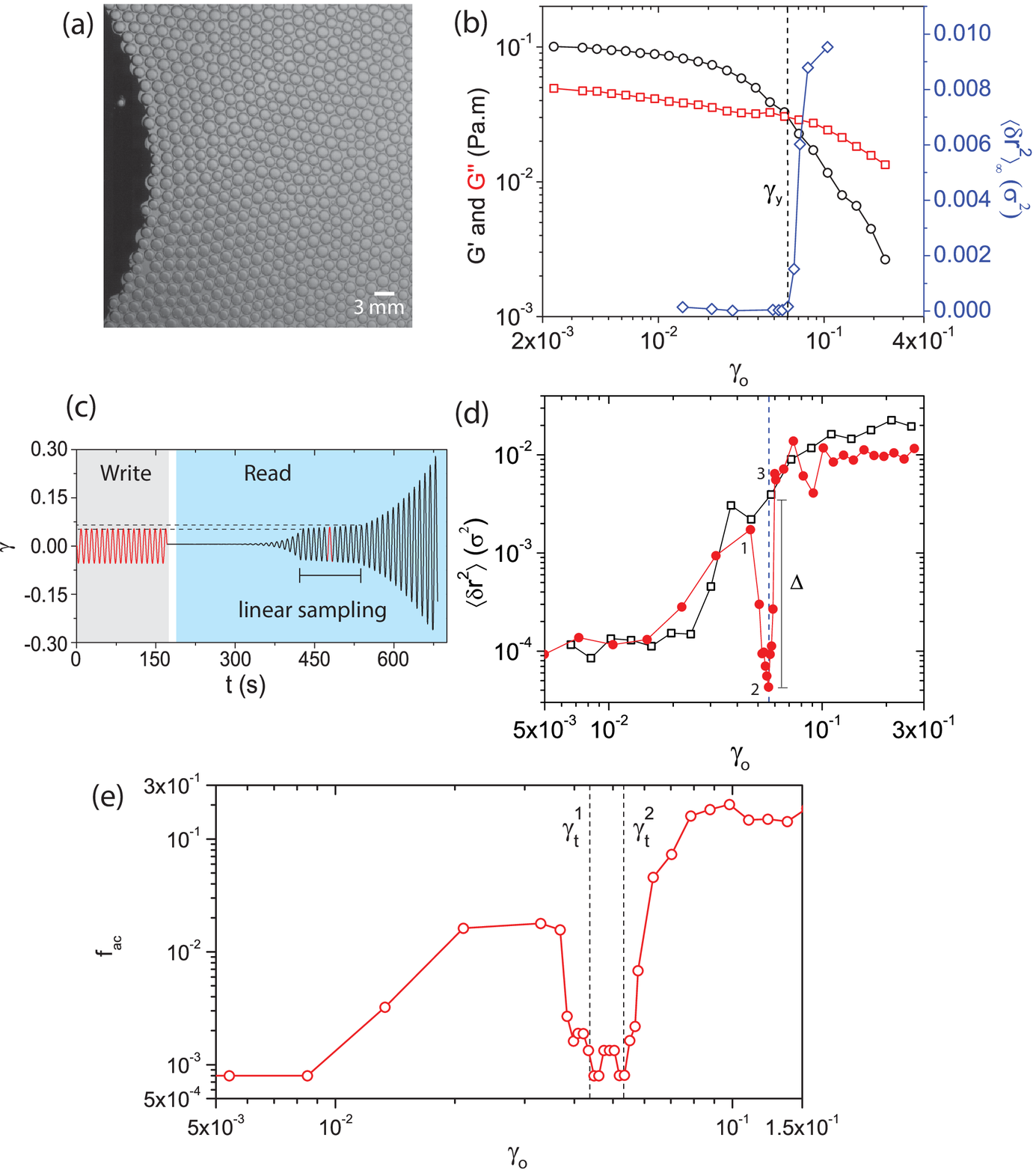}
\caption{(a) Representative image of the bubble raft. Number of small bubbles N${_S}$ to large bubbles N${_L}$ is N${_S}$/N${_L}$ $\approx$ 1.9 and average bubble diameter is $\sigma$ $\approx$ 1.25 mm. (b) Amplitude sweep measurements to quantify $\gamma_y$. $G'$(black circles), and $G''$(red squares), as a function of $\gamma_{\circ}$. Vertical line is drawn at $\gamma_{\circ} = \gamma_y$ = 0.06. ${\langle {\delta r}^2 \rangle}_{\infty}$ is shown by the blue diamonds. (c) Typical write and read protocol followed in our experiments. Data corresponds to $\gamma_t = 0.056$. Writing is done for $n = 17$ cycles. Memory was read after a 10 s pause after writing (blue shaded region). (c) Evolution of $\langle {\delta r}^2 \rangle$ with $\gamma_{\circ}$ without training (black squares) and with training at $\gamma_t$ = 0.056. (d) Multiple memories: f$_{ac}$ as a function of $\gamma_{\circ}$ during the read showing two drops corresponding to the memories of $\gamma_1$ = 0.042 and $\gamma_2$ = 0.053. Multiple memories were more evident in $f_{ac}$ as compared to $\langle {\delta r}^2 \rangle$.}
\label{Figure1}
\end{figure}

Our amorphous rafts are a mixture of small and large bubbles of diameters $\sigma_s = 1.1$ mm and $\sigma_l = 1.4$ mm, respectively, to avoid crystallization under shear (Fig. \ref{Figure1}(a) and Supplementary Material and Supplementary Fig. S1 \cite{supplementary}). The amorphous rafts were contained in a custom designed two-dimensional wide-gap circular Couette cell with an inner rotating disc of radius $R_i = 3.1$cm and an outer cylinder of radius $R_o = 7.5$cm. The inner disc was coupled to a commercial rheometer (MCR 301, Anton-Paar Austria) for applying precise mechanical forcing. High-speed imaging (Photron Fastcam SA4, Photron UK) of the rafts under shear allowed simultaneous quantification of both single-particle dynamics and rheological response. Prior to each measurement, we pre-sheared the raft at a shear rate $\dot\gamma = 50$s$^{-1}$ for 120s to avoid history effects. We first quantified the yield point of the bubble rafts by applying an oscillatory strain, $\gamma(t) = \gamma_{\circ}\sin(\omega t)$, keeping the angular frequency fixed at $\omega = 0.628$ rad/s and sweeping the strain amplitude $\gamma_\circ$. Figure \ref{Figure1}b shows the elastic and viscous  moduli, $G'$ (black circles) and $G''$ (red squares) respectively, versus $\gamma_\circ$ for the raft shown in the inset. The behavior observed is typical of a soft amorphous solid with $G'>G''$ in the linear response regime \cite{sollich1998rheological}. The onset of plasticity is characterized by the breakdown of linearity and is followed by a crossover of $G'$ and $G''$ at $\gamma_\circ = 0.06$, which we identified with $\gamma_y$.

Given that memory formation finds its footing in RIT \cite{keim2011generic}, we first confirmed the existence of this transition for our bubble rafts. In these experiments, the rafts were subjected to repeated strain oscillations $\gamma(t) = \gamma_t\sin(\omega t)$ where, $\omega = 0.628$ rad/s (see Supplemental Movie S1 \cite{supplementary}) and we simultaneously followed the change in irreversibility in the system with oscillation cycle number, $n$. We quantified irreversibility by calculating the variance in particle positions, $\left\langle {\delta {r}^2}\right\rangle = {1\over N}\sum\limits_{i = 1}^{N}\delta {r_i}^2$, between snapshots pertaining to the beginning and end of a strain cycle. Here, $\langle\rangle$ denotes an average over all particles, $\delta {r_i}^2 = (r_i(n) - r_i(n-1))^2$ with $r_i(n-1)$ and $r_i(n)$ being the initial and final positions of particle $i$ at the start and end of the $n^{th}$ oscillation cycle (i.e. at $\gamma = 0$), respectively, and $N$ is the total number of particles in the field of view. We observed that for $\gamma_t\leq\gamma_y$, $\left\langle {\delta {r}^2}\right\rangle$ dropped to zero with $n$, while for $\gamma_t>\gamma_y$, $\left\langle {\delta {r}^2}\right\rangle$ plateaued at a finite value (Supplementary Fig. S2 \cite{supplementary}). The steady state value of the variance, ${\langle\delta r^2\rangle}_{\infty}$ (blue diamonds in Fig. \ref{Figure1}b), which serves as the order parameter for the transition clearly shows that the RIT for bubble rafts is centered at $\gamma_y$ and is consistent with previous experiments and simulations \cite{nagamanasa2014experimental,fiocco2013oscillatory}.

Apart from helping confirm the existence of a RIT, the experiments described above (Supplementary Fig. S2) also served the purpose of training the raft at various $\gamma_t$'s (see Supplemental Movie S2 \cite{supplementary}). Immediately after training at each $\gamma_t$, we performed a `read' which comprised of a sequence of systematically increasing oscillatory strain amplitudes, $\gamma_{\circ}$, spanning $\gamma_t$ (Fig. \ref{Figure1}c). $\gamma_{\circ}$ was sampled logarithmically far away from $\gamma_t$ and linearly in the vicinity of $\gamma_t$ to better detect memory. As a read-out of memory, we once again quantified the variance in particle positions, that were measured stroboscopically as in write, but with one minor difference. Here, $\left\langle {\delta {r}^2}\right\rangle = {1\over N}\sum\limits_{i = 1}^{N}(r_i(\gamma_\circ+\delta\gamma_{\circ}) - r_i(\gamma_\circ))^2$ where, $\delta\gamma_{\circ}$ is the increment in the read strain amplitude between successive cycles. Figure. \ref{Figure1}d shows $\left\langle{\delta {r}^2}\right\rangle$ versus $\gamma_{\circ}$ for an untrained raft (black squares) and for the same raft after being trained at $\gamma_t = 0.056$ (red circles). The data sets were smoothened using a sliding 3-point averaging procedure. In spite of substantial irreversibility being present for $\gamma_{\circ} < \gamma_t$, the raft still retains information of the training (see Supplemental Movie S3 \cite{supplementary}) with $\left\langle {\delta {r}^2}\right\rangle$ dropping by nearly two orders-of-magnitude when $\gamma_{\circ} \approx 0.056$. We have ensured that our results are not an outcome of the specific sequential read procedure followed (see Supplemental Material \cite{supplementary}). Our findings are in line with numerical studies where the trajectory of the system in the potential energy landscape during read showed a non-trivial but closed orbit only for $\gamma_{\circ} = \gamma_t$, while for $\gamma_{\circ} \neq \gamma_t$ the orbits were open resulting in a finite irreversibility \cite{fiocco2014encoding}. Information of the training can also be seen as a stress drop when $\gamma_{\circ} \approx \gamma_t$ in the bulk rheological data (see Supplemental Material \cite{supplementary}). 

The lack of an ordering of reversible states in our rafts, $\left\langle {\delta {r}^2}\right\rangle > 0$ for $\gamma_{\circ}<\gamma_t$, opens up the possibility of encoding of multiple memories without the addition of external noise (as in \cite{fiocco2014encoding}). We attempted to form such memories by training the raft at two amplitudes $\gamma_1 = 0.042$ and $\gamma_2 = 0.053$ at once. The training sequence comprised of 11 cycles of training at $\gamma_2$ and 22 cycles of training at $\gamma_1$, and the entire set was repeated twice. Figure \ref{Figure1}e shows the fraction of active particles, $f_{ac}$, versus $\gamma_{\circ}$ during the read. Particles were denoted active if $\sqrt{\delta {r}^2_i} > 0.1\sigma$ with $\sigma = {{\sigma_s +\sigma_l}\over 2}$. Memory of both the training amplitudes is clearly evident. Taken together, these observations constitute the first experimental evidence of both single and multiple memories in amorphous solids.

\begin{figure}[tbp]
\includegraphics[width=0.8\textwidth]{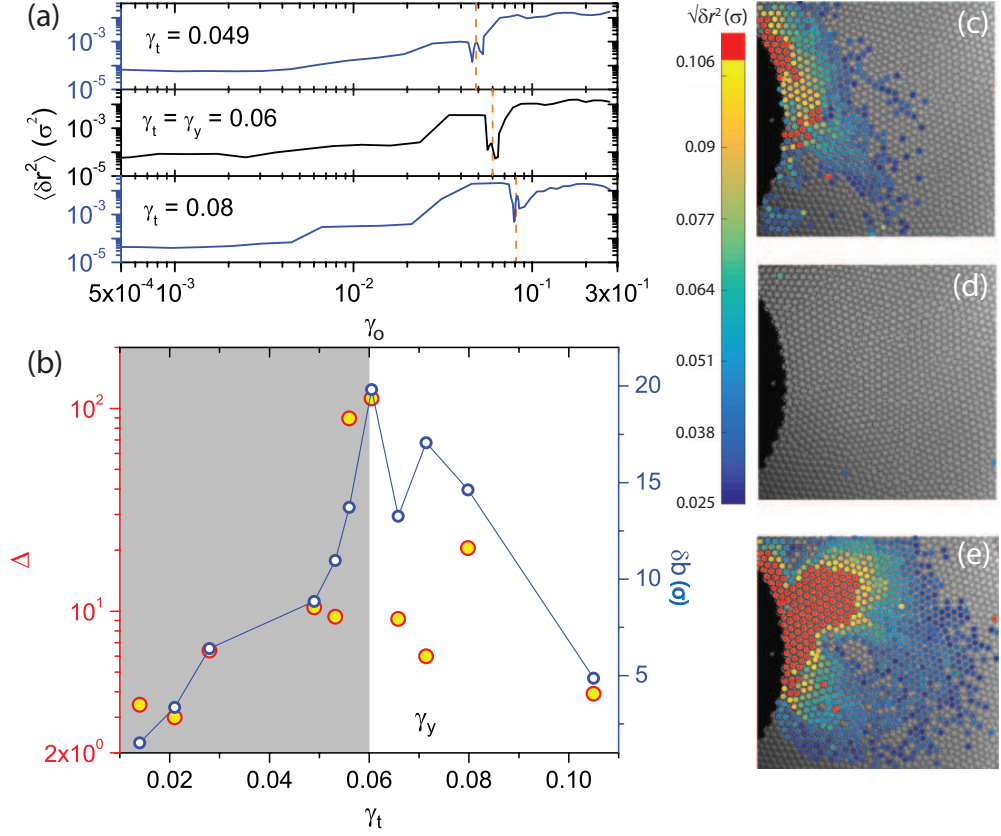}
\caption{(a) Representative read profiles for $\gamma_t$s across $\gamma_y$. (b) $\Delta$ denoted in Fig. \ref{Figure1}d as a function of $\gamma_t$ (filled red circles). $\delta b$ is shown by hollow blue circles. Grey shaded region corresponds to the pre-yield regime.(d)-(e) show stroboscopic images of the raft during the read corresponding for points labeled 1 ($\gamma_{\circ}$ = 0.046), 2 ($\gamma_{\circ}$ = 0.056) and 3 ($\gamma_{\circ}$ = 0.06) in Fig. \ref{Figure1}d, respectively. Particles are color-coded according to the magnitudes of their displacements (see color bar).}
\label{Figure2}
\end{figure}

We next turned our attention to quantifying the formation of single memories for various $\gamma_t$s spanning $\gamma_y$. Figure \ref{Figure2}a shows read profiles for a few representative $\gamma_t$s. Two features in Fig. \ref{Figure2}a stand out. First, we clearly observe memory for $\gamma_t>\gamma_y$ (vertical dashed lines in Fig. \ref{Figure2}a) during sequential reading. This is not entirely surprising given that even under overdriving ($\gamma_t>\gamma_y$), although the system settles down to a fluctuating steady state, there is still a substantial drop in irreversibility during the initial few cycles of training, leaving atleast a partial imprint of $\gamma_t$ (see Supplementary Fig. S2 \cite{supplementary}). The second and more striking feature is that the magnitude of the drop in $\left\langle {\delta {r}^2}\right\rangle$ in the vicinity of $\gamma_t$, which is simply a measure of how well the system retains information of the training, is largest for $\gamma_t = \gamma_y$. We parametrized this strength of memory by $\Delta$, which we define as ratio of $\left\langle \delta {r^2}\right\rangle$s of the untrained raft and the trained one at $\gamma_\circ = \gamma_t$ (Fig. \ref{Figure1}d). Figure \ref{Figure2}b shows $\Delta$ (filled circles) as function of $\gamma_t$ for all training amplitudes investigated. In spite of some scatter being present, the non-monotonic evolution in $\Delta$ with $\gamma_t$ with a maximum at $\gamma_y$ is indisputable. Further, $\Delta$ for $\gamma_t=\gamma_y$ is nearly two orders of magnitude larger than its corresponding values at the lowest and highest $\gamma_t$s investigated.

We gleaned further insights into the observed maximum in $\Delta$, by quantifying the spatial distribution of irreversible particles during read for various $\gamma_t$'s. Figure \ref{Figure2}c-e shows the stroboscopic images of the raft corresponding to read strains labeled 1, 2 and 3 in Fig. \ref{Figure1}d, with the particles color-coded according to the magnitudes of their displacements. For $\gamma_\circ$s straddling $\gamma_t$ (labeled 1 and 3), particles that underwent substantial irreversible displacement form a reasonably well-defined band adjacent to the inner rotating disc. Supplementary Movie S4 \cite{supplementary} shows that as $\gamma$ is increased towards $\gamma_t$, the spatial extent over which particles underwent irreversible displacements grows radially outwards from the inner rotating disc, collapses when $\gamma_{\circ} \approx \gamma_t$ and grows radially outwards again as $\gamma_{\circ}$ is further increased. To check whether this feature is related to the observed trend in $\Delta$, we measured the distance the edge of the activity field moves, $\delta b$ for various $\gamma_t$s during the read. Here, $\delta b$ is the difference in the maximum spatial extent where active particles are found, between points labeled 1 and 2 in Fig. \ref{Figure1}d. We observed a maximum in $\delta b$ in the vicinity of $\gamma_y$ (blue circles in Fig. \ref{Figure2}b) with its overall trend mimicking the one observed in $\Delta$. 

\begin{figure}[tbp]
\includegraphics[width=0.8\textwidth]{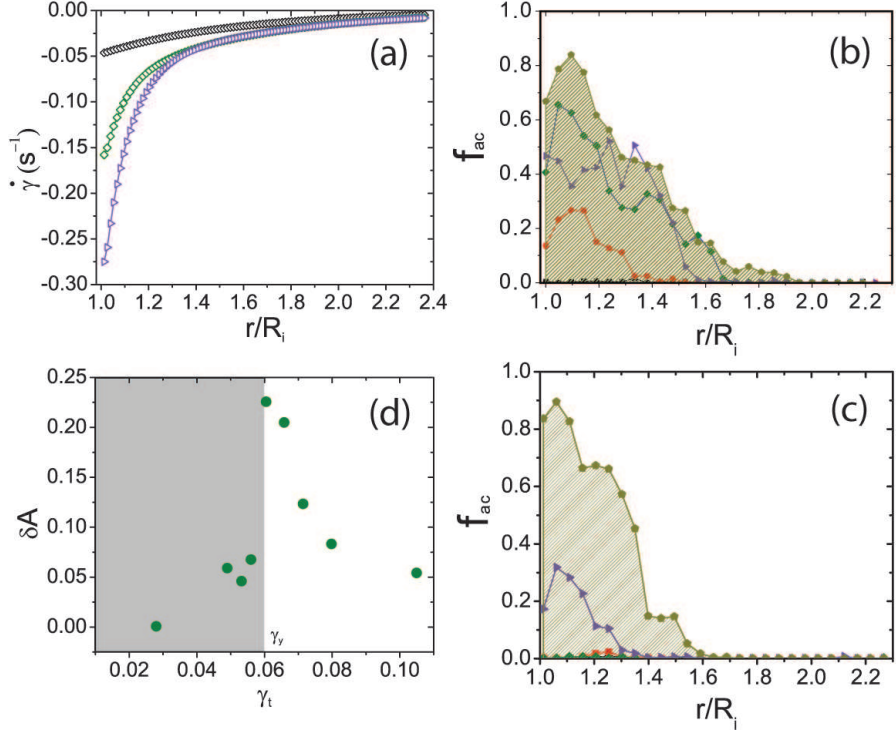}
\caption{(a) Strain rate for $\gamma_t = 0.028$ (black diamonds), $\gamma_t =\gamma_y = 0.060$ (green diamonds), and $\gamma_t = 0.071$ (right triangles). (b) and (c) $f_{ac}$ as a function of $r/R_i$ for the 2$^{\text{nd}}$ and the 16$^{\text{th}}$ training cycles, respectively. $\gamma_t = 0.028$ (black diamonds), $\gamma_t = 0.049$ (red circles), $\gamma_t =\gamma_y = 0.060$ (green diamonds), $\gamma_t = 0.071$ (violet right triangles) and $\gamma_t = 0.105$ (pale green pentagons). (d) $\delta A$ versus $\gamma_t$. $\delta A$ is dimensionless since both the abscissa and ordinate of Fig. 3b-c are dimensionless. $\delta A$ for $\gamma_t$ = 0.014 and $\gamma_t$ = 0.021 have not been shown due to difficulties in finding irreversible particles ($\sqrt{\delta {r}^2_i} > 0.1\sigma$) for these $\gamma_t$(s) during write.}
\label{Figure3}
\end{figure}

\textit{Why is there a maximum in $\Delta$ at $\gamma_t =\gamma_y$?} The extent to which the system retains memory of the input depends on how well this information was encoded and the answers therefore have to come from the write phase. Supplementary Movie S5 \cite{supplementary} shows the spatial evolution of irreversibility, when seen stroboscopically, during training at $\gamma_t$ = 0.056 $\approx \gamma_y$. We clearly see that the magnitude of particle irreversibility decays radially from the inner rotating disc and with increasing $n$, the radial extent of this activity field diminishes and vanishes completely by the end of training. Our wide-gap Couette geometry results in a stress inhomogeneity across the gap that decays as $1/r^{2}$ at a radial distance $r$ from the center of the inner moving plate \cite{katgert2010couette,lauridsen2004velocity,schall2010shear}. This stress inhomogeneity results in a curvilinear strain-rate, $\dot{\gamma}$, profile across the gap. To calculate $\dot{\gamma}(r)$ we first divided the gap into rings of width $1.22\sigma$, concentric with the inner disc, and then computed the average azimuthal velocity $v$ of the particles within each ring from subsequent images. The velocity profiles $v(r)$ were then averaged over the first ten image pairs, corresponding to 0.17s, of the strain oscillation cycle wherein the acceleration of the inner disc is practically zero. Supplementary Fig. S6 \cite{supplementary} shows the velocity profiles for various $\gamma_t$s. The velocity profiles were then smoothened to calculate $\dot{\gamma}(r) = \frac{dv}{dr}-\frac{v}{r}$. Figure \ref{Figure3}a shows $\dot{\gamma}(r)$ versus $r/R_i$ for three representative $\gamma_t$s spanning yield. A decay in the magnitude of $\dot{\gamma}(r)$ with $r$ implies that regions of the raft closer to the inner disc will take larger cycles to self-organize then regions that are farther away. Thus during training, the edge of the activity field must migrate towards the inner rotating disc. 

We next quantified the spatial evolution of activity during training by measuring $f_{ac}$ within each ring. Figure \ref{Figure3}b and c show $f_{ac}$ versus $r/R_i$, for various $\gamma_t$s at the beginning and end of the training, respectively. When $\gamma_t<<\gamma_y$, the strain amplitude is too weak to cause substantial irreversible rearrangements and the final particle packing at the end of training is not too different from the one at the start (red squares in Fig. \ref{Figure3}b and c). Since a sizeable fraction of the system is unable to reconfigure, $\Delta$ is small. In the $\gamma_t >> \gamma_y$ regime, the strain amplitude is large enough to cause considerable irreversible rearrangements at the beginning of the training but significant irreversibility also remains at the end, i.e. in the fluctuating steady state (pale green symbols in Fig. \ref{Figure3}b and c). Thus, $\Delta$ is once again small. For $\gamma_t \approx \gamma_y$, the edge of the activity field sweeps the largest area between the beginning and end of the training process, resulting in maximal reconfiguration of the system and hence $\Delta$ is large (olive-green diamonds in Fig. \ref{Figure3}b and c). The area swept between the beginning and end of training $\delta A$ for various $\gamma_t$s is shown in (Fig. \ref{Figure3}d). Like $\Delta$, $\delta A$ is non-monotonic and is maximal in the vicinity of $\gamma_y$. A previous study on amorphous bubble rafts under nearly identical experimental conditions found a `flow cooperativity length', finite only in the jammed state, which results in strong non-local effects during the flow of these rafts \cite{katgert2010couette}. These non-local effects are compounded by the presence of a stress inhomogeneity as in the present study \cite{goyon2008spatial}. Furthermore, studies on colloidal glasses under oscillatory shear have shown that spatially cooperative relaxation dynamics is maximal in the vicinity of the yielding transition \cite{nagamanasa2014experimental}. Whether such spatial correlations have a direct bearing on the observed maximum in $\Delta$ remains to be seen.

\begin{figure}[tbp]
\includegraphics[width=0.8\textwidth]{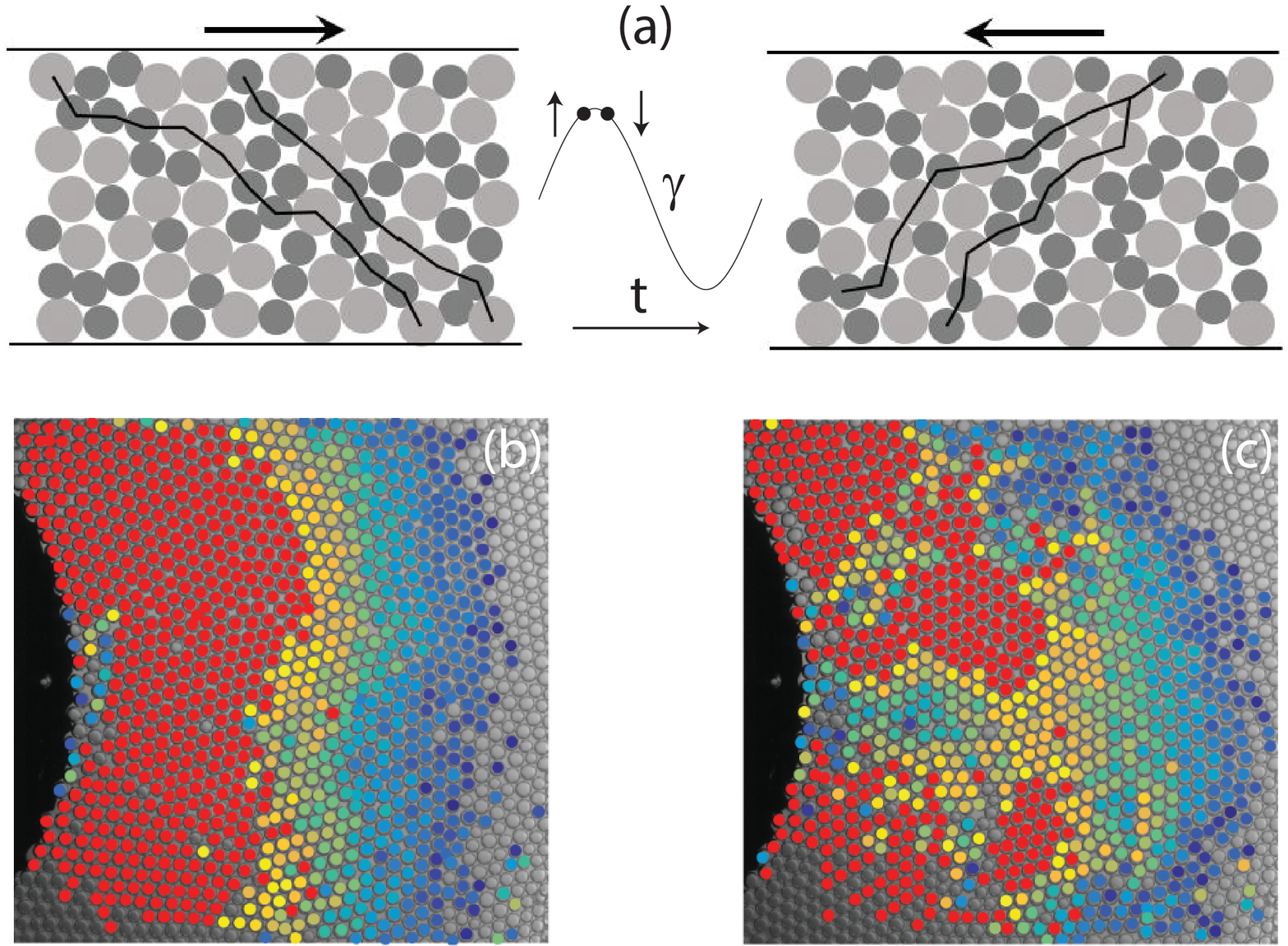}
\caption{(a) Schematic of force chains in jammed packings immediately after strain reversal. Although the configuration is practically unchanged immediately after reversal, the contact network is not. (b) Particle displacement map for $\gamma_{\circ}=0.073$ during read. This strain corresponds to the maximum before the drop (equivalent of point labeled 1) for a $\gamma_t =0.079$. (c) Particle displacement map after the first cycle of write for $\gamma_t =0.071$ which is closest to point 1 in our study.}
\label{Figure4}
\end{figure}

We finally return to our observation of a finite irreversibility for $\gamma_{\circ}<\gamma_t$ during read (red circles in Fig. \ref{Figure1}d). Unlike memory formation in non-Brownian suspensions where adding noise results in finite particle displacements and results in $\left\langle {\delta {r}^2}\right\rangle > 0$ for $\gamma_{\circ}<\gamma_t$ during read \cite{keim2011generic,paulsen2014multiple}, in jammed packings like ours, it remains unclear. Below we provide a plausible explanation. Earlier studies on cyclically sheared dense packings have found that the fraction of contacts broken between nearest-neighbors, $f_b$, peaks during strain reversal \cite{slotterback2012onset}. On training at a given $\gamma_t$ however, $f_b$ drops and reaches a steady state and this suggests that the same links are broken during subsequent strain cycles. The contact network, however, which points predominantly along the compression axis, cannot remain identical immediately after strain reversal (Fig. \ref{Figure4}a) \cite{archive}. Particle configuration therefore retraces a different path after strain reversal, albeit it closes in on itself after a full cycle for $\gamma_{\circ}=\gamma_t$ and is consistent $\left\langle {\delta {r}^2}\right\rangle \approx 0$ \cite{fiocco2014encoding}. $f_b$ and also the contact network, for any $\gamma_{\circ}\neq\gamma_t$ during read, will not have evolved to a steady state \cite{slotterback2012onset} and the particle packing should essentially behave like an untrained one (open orbits and $\left\langle {\delta {r}^2}\right\rangle > 0$). This naive expectation is indeed borne out by our observations where the $\left\langle {\delta {r}^2}\right\rangle$ for the untrained raft (black symbols) almost follows the trained one (red symbols) up until the point labeled 1 in Fig. \ref{Figure1}d. Figure \ref{Figure4}b, shows the particle displacement map during read at $\gamma_{\circ}=0.073$ which is the equivalent of point labeled 1 for $\gamma_t =0.079$. The spatial extent of irreversible displacements is nearly identical to that after the first cycle of write for $\gamma_t = 0.071$ (Fig. \ref{Figure4}c), which is the $\gamma_t$ closest to $\gamma_{\circ}=0.073$ in our study. Perhaps the most intriguing finding, which our study does not provide answers to, is how the raft manages to retain memory of $\gamma_t$ in spite of behaving like an untrained one close to $\gamma_t$.

Collectively, our study provides the first direct experimental evidence of both single and multiple mechanical memories in athermal amorphous solids under cyclic shear. Remarkably, the strength of these single memories is maximal near $\gamma_y$ and is intimately connected to the extent to which particle irreversibility spatially evolves during the encoding process. These findings suggest that the recent observations of growing spatial correlations in the vicinity of the yielding transition \cite{nagamanasa2014experimental,ghosh2017direct} may have a direct role in the formation of mechanical memories. Given that numerical studies find cyclic shear to be a potential route to prepare well-annealed glasses \cite{leishangthem2017yielding}, it is tempting to wonder if the strength of memory formation can in fact be used as readout for ultrastability. A natural step forward would be to explore connections between memory formation and the evolution of force networks with training. We believe that frictionless athermal systems like dense assemblies of bubble rafts and emulsions, where force chains can be quantified through shape distortions \cite{zhou2006measurement}, will prove to be an ideal platform for these measurements.

\begin{references}

\bibitem{paulsen2014multiple}
J.D. Paulsen, N.C. Keim, and S.R. Nagel, Phys. Rev. Lett. {\bf{113}}, 068301 (2014).
\bibitem{keim2011generic}
N.C. Keim, and S.R. Nagel, Phys. Rev. Lett. {\bf{107}}, 010603 (2011).
\bibitem{povinelli1999noise}
M.L. Povinelli, S.N. Coppersmith, L.P. Kadanoff, S.R. Nagel, and S.C. Venkataramani, Phys. Rev. B {\bf{59}}, 4970 (1999).
\bibitem{jonason1998memory}
K. Jonason, E. Vincent, J. Hammann, J.P. Bouchaud, and P. Nordblad, P, Phys. Rev. Lett. {\bf{81}}, 3243 (1998).
\bibitem{sethna1993hysteresis}
J.P. Sethna, K. Dahmen, S. Kartha, J.A. Krumhansl, B.W. Roberts, and J.D. Shore, Phys. Rev. Lett. {\bf{70}}, 3347 (1993). 
\bibitem{fiocco2014encoding}
D. Fiocco, G. Foffi, and S. Sastry, Phys. Rev. Lett. {\bf{112}}, 025702 (2014).
\bibitem{pine2005chaos}
D.J. Pine, J.P. Gollub, J.F. Brady, and A.M. Leshansky, Nature {\bf{438}}, 997 (2005).
\bibitem{corte2008random}
L. Corte, P.M. Chaikin, J.P. Gollub, and D.J. Pine, Nat. Phys. {\bf{4}}, 420 (2008). 
\bibitem{fiocco2013oscillatory}
D. Fiocco, G. Foffi, and S. Sastry, Phys. Rev. E {\bf{88}}, 020301 (2013).
\bibitem{regev2013onset}
I. Regev, T. Lookman, and C. Reichhardt, Phys. Rev. E {\bf{88}}, 062401 (2013).
\bibitem{nagamanasa2014experimental}
K.H. Nagamanasa, S. Gokhale, A.K. Sood, and R. Ganapathy, Phys. Rev. E {\bf{89}}, 062308 (2014).
\bibitem{regev2015reversibility}
I. Regev, J. Weber, and C. Reichhardt, K.A. Dahmen, and T. Lookman, Nat. Commun. {\bf{6}}, 8805 (2015); N.C. Keim, and P.E. Arratia, Soft Matter {\bf{9}}, 6222 (2013); E.D. Knowlton, D.J. Pine, and L. Cipelletti, Soft Matter {\bf{10}}, 6931 (2014).
\bibitem{keim2013multiple}
N.C. Keim, J.D. Paulsen, and S.R. Nagel, Phys. Rev. E {\bf{88}}, 032306 (2013).
\bibitem{goyon2008spatial}
J. Goyon, A. Colin, G. Ovarlez, A. Ajdari, and L. Bocquet, Nature {\bf{454}}, 84 (2008). 
\bibitem{katgert2010couette}
G. Katgert, B.P. Tighe, M.E. M{\"o}bius and M. van Hecke, Europhysics Letters {\bf{90}}, 54002 (2010).
\bibitem{schall2007structural}
P. Schall, D.A. Weitz, and F. Spaepen, Science {\bf{318}}, 1895 (2007).
\bibitem{chikkadi2011long}
V. Chikkadi, G. Wegdam, D. Bonn, B. Nienhuis, and P. Schall, Phys. Rev. Lett. {\bf{107}}, 198303 (2011).
\bibitem{ghosh2017direct}
A. Ghosh, Z. Budrikis, V. Chikkadi, A.L. Sellerio, S. Zapperi, and P. Schall, Phys. Rev. Lett. {\bf{118}}, 148001 (2017). 
\bibitem{shi1982potential}
L.T. Shi, and A.S. Argon, Philosophical Magazine A {\bf{46}}, 255 (1982).
\bibitem{bragg1947dynamical}
W.L. Bragg, and J.F. Nye, Proc. R. Soc. Lond. A {\bf{190}}, 474 (1947).
\bibitem{debregeas2001deformation}
G. Debregeas, H. Tabuteau, and J.M. Di Meglio, Phys. Rev. Lett. {\bf{87}}, 178305 (2001).
\bibitem{lauridsen2004velocity}
J. Lauridsen, G. Chanan, and M. Dennin, Phys. Rev. Lett. {\bf{93}}, 018303 (2004).
\bibitem{schall2010shear}
P. Schall, and M. van Hecke, Annual Review of Fluid Mechanics {\bf{42}}, (2010).
\bibitem{supplementary} 
See Supplemental Material 
\bibitem{sollich1998rheological}
P. Sollich, Phys. Rev. E {\bf{58}}, 738 (1998).
\bibitem{slotterback2012onset} 
S. Slotterback, M. Mailman, K. Ronaszegi, M. Van Hecke, M. Girvan, and W. Losert, Phys. Rev. E {\bf{85}}, 021309 (2012).
\bibitem{archive}
M.L. Falk, M. Toiya, and W. Losert, arXiv preprint arXiv:0802.1752v2, (2010). 
\bibitem{leishangthem2017yielding}
P. Leishangthem, A.D.S. Parmar, and S. Sastry, Nat. Commun. {\bf{8}}, 14653 (2017).
\bibitem{zhou2006measurement}
J. Zhou, S. Long, Q. Wang, and A.D. Dinsmore, Science {\bf{312}}, 1631 (2006).
	
\end{references}
\end{document}